\documentclass[sigconf]{acmart}

\AtBeginDocument{%
  }

\copyrightyear{2026}
\acmYear{2026}
\setcopyright{cc}
\setcctype{by}
\acmConference[UIST '26]{The 39th Annual ACM Symposium on User Interface Software and Technology}{November 02--05, 2026}{Detroit, MI, USA}
\acmBooktitle{The 39th Annual ACM Symposium on User Interface Software and Technology (UIST '26), November 02--05, 2026, Detroit, MI, USA}
\acmDOI{10.1145/3830398.3830593}
\acmISBN{979-8-4007-2856-3/2026/11}

\acmSubmissionID{3890}
\usepackage[nameinlink,capitalise]{cleveref}
\usepackage{booktabs}
\usepackage{float}

\newcommand{\subcref}[2]{\hyperref[#1]{\cref*{#1}#2}}
\crefname{figure}{Fig.}{Figs.}
\Crefname{figure}{Figure}{Figures}
\crefname{table}{Tab.}{Tabs.}
\Crefname{table}{Table}{Tables}
\crefname{section}{Sec.}{Secs.}
\Crefname{section}{Section}{Sections}
\usepackage{enumitem}

\newcommand{\rc}[1]{#1}
\newcommand{\rw}[1]{#1}

\usepackage{xspace}

\begin{document}

\title{RegulAR: Graph-Grounded Error Recognition and Assistance for Procedural Tasks in Augmented Reality}

\author{Yi-Lin Ye}
\orcid{0009-0000-9986-4420}
\affiliation{
  \institution{The Hong Kong University of Science and Technology}
  \city{Hong Kong SAR}
  \country{China}
}
\email{yyeaz@connect.ust.hk}

\author{Jindu Wang}
\orcid{0009-0009-4028-4662}
\affiliation{
  \institution{The Hong Kong University of Science and Technology}
  \city{Hong Kong SAR}
  \country{China}
}
\email{jwangki@connect.ust.hk}

\author{Hiu Tung Wong}
\orcid{0009-0005-7330-099X}
\affiliation{
  \institution{The Hong Kong University of Science and Technology}
  \city{Hong Kong SAR}
  \country{China}
}
\email{htwongbe@connect.ust.hk}

\author{Shuchang Xu}
\orcid{0000-0002-7642-9044}
\affiliation{
  \institution{The Hong Kong University of Science and Technology}
  \city{Hong Kong SAR}
  \country{China}
}
\email{sxuby@connect.ust.hk}

\author{Huamin Qu}
\orcid{0000-0002-3344-9694}
\affiliation{
  \institution{The Hong Kong University of Science and Technology}
  \city{Hong Kong SAR}
  \country{China}
}
\email{huamin@ust.hk}

\author{Wong Kam-Kwai}
\authornote{Corresponding author.}
\orcid{0000-0002-2813-1972}
\affiliation{
  \institution{The Hong Kong University of Science and Technology}
  \city{Hong Kong SAR}
  \country{China}
}
\email{kkwongar@connect.ust.hk}

\renewcommand{\shortauthors}{Ye et al.}

\begin{abstract}
Errors are inevitable in procedural tasks, yet most AR guidance systems focus on step-by-step instruction delivery rather than helping users recognize and recover from mistakes.
We present \textit{RegulAR}, an AR task assistant for procedural error recognition and recovery.
\textit{RegulAR} models task instructions as a hierarchical dependency graph and combines this structure with a Multimodal Large Language Model (MLLM) to interpret egocentric observations during execution.
This enables \textit{RegulAR} to track progress, identify deviations by error type, estimate their impact on later steps, and deliver appropriately salient interventions through an in-situ head-up display that visualizes task state and recovery guidance.
By making procedural structure explicit, \textit{RegulAR} supports not only next-step guidance, but also reasoning about what went wrong, why it matters, and how users can get back on track.
In a within-subject study ($N=12$), \rw{participants reported better task-structure understanding and recovery support with \textit{RegulAR}} than the MLLM-only baseline.
\end{abstract}



\begin{teaserfigure}
  \includegraphics[width=\textwidth]{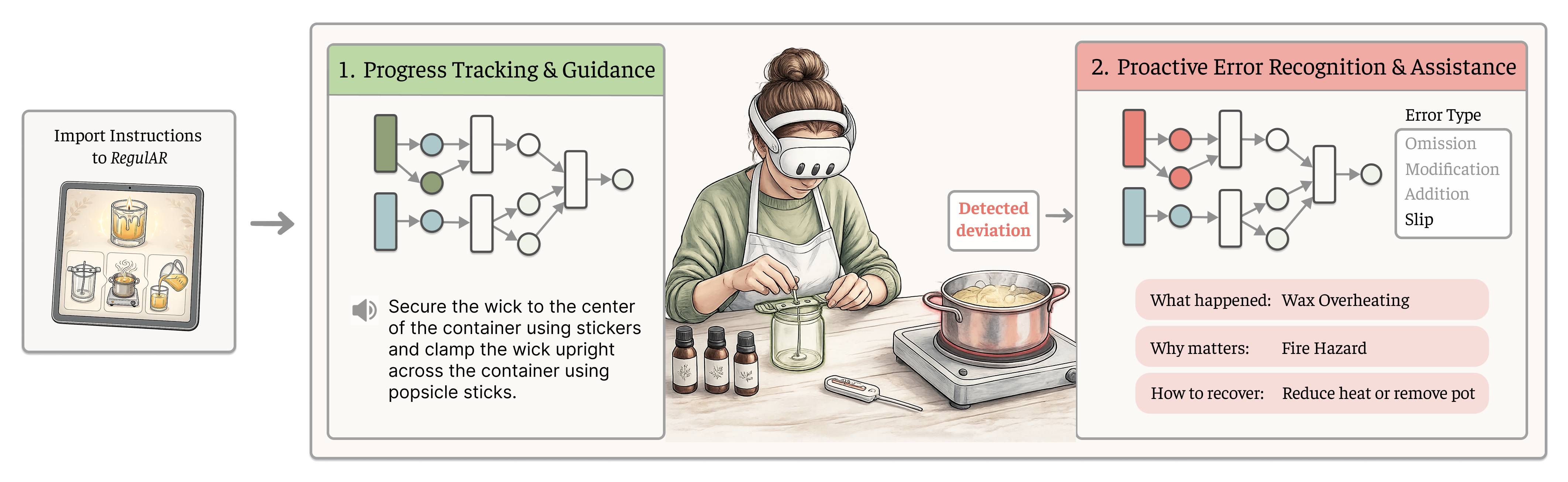}
  \caption{\textit{RegulAR} is an AR assistant for proactive error-aware procedural guidance. A user imports a candle tutorial, from which \textit{RegulAR} generates a task dependency graph. As the user proceeds, \textit{RegulAR} tracks progress and provides context-aware guidance. While detecting errors, \textit{RegulAR} highlights the affected action and provides recovery assistance. (Credit: Gemini 3)}
  \Description{A usage scenario of RegulAR during a candle-making task. A user imports an online tutorial, which RegulAR converts into a task dependency graph. During execution, the system tracks progress and provides guidance while the user secures the wick. At the same time, RegulAR detects an error that wax overheats, highlights the affected step, and provides assistance explaining what happened, why it matters, and how to recover.}
  \label{fig:teaser}
\end{teaserfigure}


\maketitle

\section{Introduction}
Procedural tasks require interacting with physical objects in sequences governed by step dependencies, ordering constraints, and evolving object states~\cite{henderson_procedural_2011}, spanning settings from cooking~\cite{zhai_cook_2020} and mechanical assembly~\cite{liu_instrumentar_2023, zhao_guidedreality_2025} to laboratory procedures~\cite{zhou_chemical_2024}. The challenge extends beyond remembering the next step: users must also judge whether the current task state is on track, whether prerequisites have been satisfied, whether actions have been completed correctly, and whether execution is safe to continue. This demand follows from two properties of procedural work. Human action is fallible~\cite{reason_bmj2000,reason_rstb1990}, so deviations are expected even in familiar tasks, yet users are physically and cognitively occupied with execution and have little capacity to notice them. Procedural tasks are also dependency-structured, so a local mistake can change which later actions remain valid~\cite{henderson_procedural_2011,li_egoproce_2026}, turning a minor deviation into cascading consequences. Expecting users to diagnose deviations and plan recovery on their own places too much burden at the moment when assistance is most needed.

Augmented Reality (AR) is well-suited for procedural assistance: it can present information in situ while work unfolds in the physical environment, reducing context switching and supporting hands-busy execution~\cite{tang_chi2003,henderson_ARRepair_2011,adaptutar_chi2021}. Prior AR systems have demonstrated this for instruction delivery and progress verification~\cite{tutoriallens_sui2021,liu_instrumentar_2023, instructar_uistadj2024}, and recent MLLM-powered assistants further enable real-time interpretation of visual observations and situated question answering~\cite{castelo_argus_2024,huh_vid2coach_2025,huang_vinci_2025,yang_egolife_2025}. Yet these systems are designed around nominal execution and provide little support once users deviate~\cite{henderson_ARRepair_2011}. Current assistants remain query-driven, and recent video-based error recognition work~\cite{lee_gtg_2025,lee_egoped_2024}, while improving deviation detection, does not provide proactive recovery support in a real-time AR setting. Users are still left to do the hardest part themselves: noticing that a deviation matters, understanding its consequences, and figuring out how to recover.

To understand what support users actually need when execution deviates, we conducted a \rw{needs-finding study} ($N=6$) with reactive MLLM-based assistants. Participants were not asking \rw{the assistant} for the next step; they were asking whether their current state was still acceptable and what a deviation would affect. This revealed three challenges. First, many errors are relational. An action can look plausible in isolation, yet violate a hidden prerequisite or threaten \rw{later} progress. Second, real-time egocentric interpretation is inherently ambiguous. Observations are partial and \rw{unfold over time}, and MLLMs struggle to maintain logical and temporal dependencies in streaming settings~\cite{li_egoproce_2026}. Third, interventions must be timely without becoming overbearing. Weak intervention allows errors to cascade, while aggressive interruption reduces user control~\cite{iqbal_bailey_chi2008,proactivity_dilemma_cui2022}. These findings motivate our research question: \emph{How can AR assistants proactively detect, diagnose, and support recovery from errors during real-time procedural task execution?}

To answer this, we present \textit{RegulAR}, an AR task assistant for procedural error detection and recovery. \textit{RegulAR} represents task instructions as a hierarchical dependency graph that captures subgoals, prerequisite relations, and valid execution paths, and combines this with MLLM-based egocentric interpretation to track progress, detect four categories of active errors, and estimate downstream impact in real time.
Guided by an active-error framing inspired by Reason's Swiss cheese model~\cite{reason_bmj2000, reason_rstb1990} and four error \mbox{categories} from prior work~\cite{lee_egoped_2024},
\textit{RegulAR} structures feedback around \emph{what} happened, \emph{why} it matters, and \emph{how} to recover, presented through an in-situ HUD that visualizes the task state, affected steps, and recovery cues directly in the user's view.

Our contributions are threefold:
\begin{itemize}[leftmargin=*, topsep=0pt]
    \item We position procedural error detection and recovery as a core design target for AR task assistance, and through a \rw{needs-finding} study and an active-error framing, identify four error categories and derive design requirements for proactive, situated support.
    \item We present \textit{RegulAR}, which combines a hierarchical task dependency graph, MLLM-based egocentric reasoning, and an in-situ HUD to detect deviations, estimate downstream impact, and deliver layered recovery guidance during real-world task execution.
    \item Through a log-based technical evaluation and a within-subject user study ($N=12$), we \rw{provide preliminary evidence that}
    \textit{RegulAR} \rw{supports} procedural state tracking, error detection, task-structure understanding, and perceived error support compared to a \rw{prompt-only MLLM baseline}, while surfacing trade-offs between intervention timing and user autonomy.
\end{itemize}

\section{Related Work}
This section reviews prior work on (1) AR-based task guidance for supporting procedural execution, and (2) computational approaches to detecting errors in procedural tasks. 
\subsection{AR Task Guidance}
\rw{AR task guidance has been widely studied to support hands-busy execution of procedural tasks.
Existing work has explored different approaches to presenting procedural guidance within the physical environment, including authoring in-situ AR instructions~\cite{subramanian_processar_2021} and overlaying step-by-step instructions directly onto physical scenes~\cite{liu_instrumentar_2023}, reducing context switching compared to external manuals.}
\rw{These approaches have since} been extended to a range of domains, including everyday activities such as cooking~\cite{zheng_aroma_2025,li_oscar_2025} and industrial maintenance~\cite{xue_arassistedavionics_2024}, where guidance is tightly coupled with physical interaction.
To improve usability under spatial and attentional constraints, subsequent work explored different strategies for information presentation. These include simplifying textual instructions~\cite{wu_artist_2024}, \rw{anchoring information} to physical objects or locations~\cite{chen_papertoplace_2023, Wong2026LandSAR, Kentaro2025TangibleNet, Zhang2026ArtiVision}, and adapting content placement to user viewpoint and occlusion~\cite{castelo_argus_2024,zhao_guidedreality_2025,li_situationadapt_2024, lu2025egoexo}. Multimodal delivery, such as coordinating visual overlays with audio guidance, has further been used to balance attention and reduce disruption during task execution~\cite{cho_evalxrmessagenotification_2025,cho_persistentassistant_2025}.
\rc{Collectively, these approaches improve how procedural instructions are presented to users. However, they largely assume a linear execution flow and provide limited support for reasoning about task progression beyond the current step.}

Recent systems incorporate Large Language Models (LLMs) and MLLMs to enable adaptive and mixed-initiative guidance. By leveraging egocentric perception, these systems can infer user intent, anticipate next actions, and provide context-aware feedback~\cite{li_omniaction_24,cai_aiget_2025}. They also support more flexible interaction paradigms, such as transforming instructional videos into interactive assistants~\cite{huh_vid2coach_2025} or delivering proactive feedback based on attention or working memory models~\cite{pei_attentionar_2025,pu_promemassist_2025,xu_memoryreviver_2024}. 
However, such systems typically \rw{reason} over immediate observations, without maintaining a persistent representation of task state.
As a result, their assistance remains short-term and observation-driven, making it difficult to track long-term progress, interpret errors, or reason about their impacts throughout the task.
Our approach addresses these limitations by representing procedural tasks as dependency graphs that explicitly capture task structure and support continuous state tracking. Combined with MLLM-based perception, this design enables proactive error recognition and structured recovery, while helping users understand task progress and dependencies during execution.

\subsection{Procedural Error Recognition and Analysis}
Detecting errors in procedural tasks is particularly challenging in egocentric settings, where continuous camera motion, partial observability, and viewpoint variability complicate reliable perception and reasoning \cite{lee_egoped_2024}.
Recent works have explored machine learning approaches for error detection and localization in egocentric videos \cite{deng_video_2023, seminara_diffgraph_2024}, but these methods require anomaly videos and labeled datasets for training \cite{qian_svip_2022,reza_data_2023, sener_assembly101_2022, peddi_captaincook4d_2024}, limiting their transferability to everyday activities and their ability to distinguish diverse error types.
To address this limitation, Lee et al. modeled procedural structure using task graphs inferred from anomaly-free videos~\cite{lee_egoped_2024, lee_gtg_2025}. 
These approaches rely on aligning full video sequences with task graphs, making them better suited for offline analysis than for real-time interactive assistance.
Recent studies have therefore explored zero-shot error recognition using MLLMs \cite{flaborea_prego_2024, zanella_harnessing_2024,hazra_egotv_2023}. 
For example, Flaborea et al. compared anticipated actions predicted by MLLMs with observed actions to identify errors.
Despite these advances, detected errors are rarely translated into actionable assistance. 

Effective procedural support requires reasoning not only about what went wrong, but also about how to preserve task integrity \cite{kung2025changedchangedstatechangecounterfactuals}. 
Structured representations such as task graphs and dataflow diagrams have been explored to model procedural dependencies and support planning ~\cite{mao_actiondynamicstaskgraphs_2023, yokoi_mermaidllm_2025, seminara_diffgraph_2024}. 
However, these representations are primarily used for instruction organization \cite{liu_instrumentar_2023, yang2023vid2seq}, visualization \cite{castelo_argus_2024, yokoi_mermaidllm_2025}, or offline reasoning \cite{lee_gtg_2025}, rather than to model task execution as a dynamic, stateful process during interaction. 
As a result, task correctness is often inferred implicitly from step completion, limiting current systems' ability to reason about how deviations affect subsequent steps or to support structured recovery during execution.
In this work, we present an AR-based guidance system that maintains an explicit task model for online error recognition, impact estimation, and dependency-aware recovery, enabling users to track progress and recover from errors during execution.
\section{\rw{Preliminary Needs-Finding Study}}
\rw{To understand the challenges users face when recovering from procedural breakdowns, we conducted a needs-finding study using reactive assistants. Our goal was to identify recurring user needs and design opportunities rather than to evaluate existing systems.}

\subsection{Methods}
\textbf{Participants.} We recruited six participants (P1-P6, 3 female, 3 male; $M=24.17$, $SD=1.17$) from a local university. All participants reported sufficient English proficiency and moderate familiarity with AR ($M=2.33$, $SD=0.52$) on a 5-point Likert scale \rc{(\Cref{tab:formative_participants})}.

\begin{figure}
    \centering
    \includegraphics[width=\linewidth]{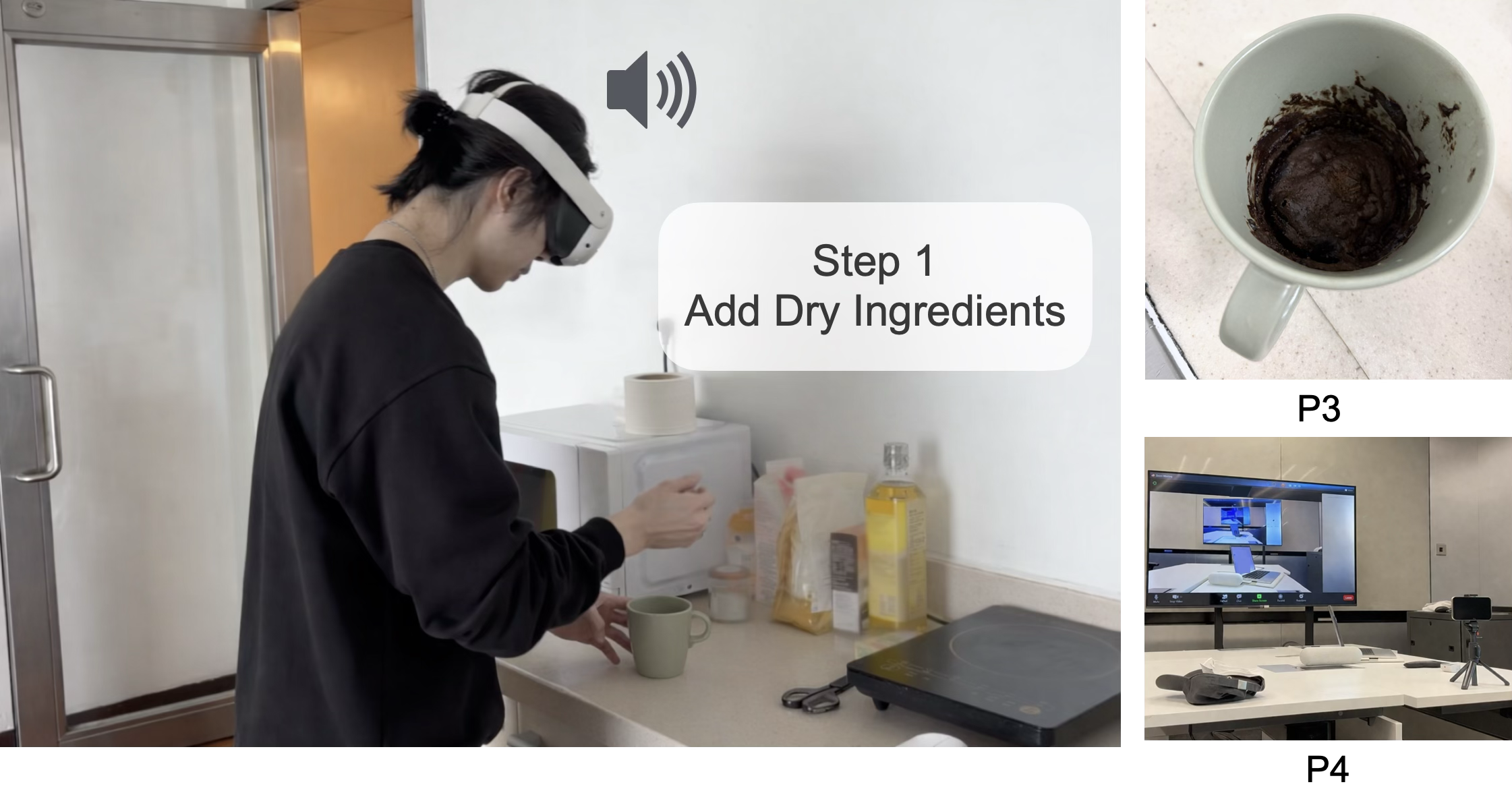}
    \caption{Settings and representative results from the \rw{needs-finding study}. Participants wore AR headsets with GPT-4V access to complete procedural tasks. Images were re-generated to remove personal information.}
    \Description{This figure shows the needs-finding study setup and representative results. A participant wears an AR headset and follows a semi-transparent instruction panel with step-related text and voice guidance while preparing ingredients. Two smaller panels show the final results of the mug cake and hybrid meeting setup tasks completed by participants P3 and P4.}
    \label{fig:formative}
\end{figure}

\textbf{Procedure.} 
After providing informed consent and receiving a study briefing, each participant completed one of two procedural tasks while wearing a Meta Quest 3 headset that displayed static AR instructions and provided optional voice access to GPT-4V in the headset camera feed (\Cref{fig:formative}).
This setup approximated a common support paradigm in which users follow instructional content while consulting a reactive AI assistant when needed.
P1-P3 performed a \textit{Hybrid Meeting Setup} task involving audio \cite{markus_hybrid_meeting_video}, display, and connectivity configuration, while P4-P6 performed a \textit{Mug Cake} task based on a professionally written recipe \cite{drummond_mugcake}. 
We selected these tasks \rw{for their dependency-sensitive, context-dependent procedures and natural error opportunities, and introduced representative failure points to elicit recovery reasoning}.
Each session concluded with a semi-structured interview on breakdowns, error-recognition strategies, recovery reasoning, and expectations for support. 
We analyzed first-person recordings, interaction logs, and interview transcripts using thematic analysis~\cite{Braun01012006Thematic}, focusing on how participants detected problems, reasoned about their consequences, and sought help during execution.

\subsection{\rw{Observations}}
\rw{We identified three recurring observations from the needs-finding study.}

\textbf{\rw{O1}: Task execution requires explicit reasoning about both local validity and global progress.}  
Participants frequently paused not to recall the next step, but to assess whether their current state was correct and whether they could proceed, \textit{“Is this right?” “Can I move on?”} (P2, P4, P6). This need emerged at two levels. At the \textit{local level}, correctness depended on implicit relationships between tools, materials, and outcomes, making state validity difficult to verify. At the \textit{global level}, participants lacked awareness of overall task structure, feasible actions, and how current actions related to downstream progress. As a result, users repeatedly reconstructed task state during execution, leading to fragmented reasoning and delayed error recognition (P1, P2).

\textbf{\rw{O2}: Errors are diverse but lack explicit interpretation support.}  
Participants encountered heterogeneous errors, including omitted steps, incorrect actions, and intentional modifications. These errors differed in both cause and consequence, yet existing instruction-based support treated them uniformly. Without explicit differentiation, participants relied on ad hoc reasoning to determine whether a deviation required correction, could be tolerated, or would affect subsequent steps (P5). This made it difficult to assess downstream impact, often resulting in inconsistent recovery strategies and delayed responses to consequential errors.

\textbf{\rw{O3}: Users expect informative feedback, but are sensitive to disruption.}  
Participants valued support in identifying and resolving issues, but reported that verbose or intrusive feedback disrupted task flow. They preferred concise problem indications, with additional explanation available on demand. This reflects a tension between informativeness and interaction overhead in hands-busy, attention-constrained settings. Without careful control, feedback either interrupts ongoing activity or fails to provide sufficient assistance for effective recovery.

\subsection{Design Requirements}
\rw{Based on these observations and prior literature, we summarize three design requirements that guided the design of \textit{RegulAR}.}

\textbf{DR1: Externalize task state to support validity and progress awareness.}
Procedural support should provide an explicit and inspectable representation of task state rather than \rw{inferring it} from step sequences.
This includes externalizing the conditions for action execution and completion, as well as exposing task structure and progress so users can understand their current position, what actions are feasible, and how local decisions affect subsequent steps.
By grounding assistance in a continuously updated state representation, the system can reduce cognitive overhead and enable more reliable reasoning about both correctness and progress.

\textbf{DR2: Differentiate errors based on type and downstream impact to guide intervention.}
Procedural support should explicitly characterize errors based on both their semantic type and their impact on \rw{downstream dependencies and task progression. By doing so,} the system can prioritize consequential errors while avoiding unnecessary intervention for minor issues. This enables more targeted, context-aware support that aligns intervention strategies with the significance of errors.

\textbf{DR3: Deliver concise, progressive assistance that adapts to error significance and user demand.}
Procedural support should adopt a progressive assistance strategy that balances clarity with minimal disruption.
Feedback should be concise by default, focusing on identifying the issue, while allowing users to access further explanation and recovery assistance on demand.
In addition, assistance should adapt to the significance of errors, ensuring that critical issues receive sufficient attention while minor ones remain unobtrusive, thereby supporting effective intervention without overwhelming users.
\section{RegulAR}
\begin{figure*}[t]
    \centering
    \includegraphics[width=\textwidth]{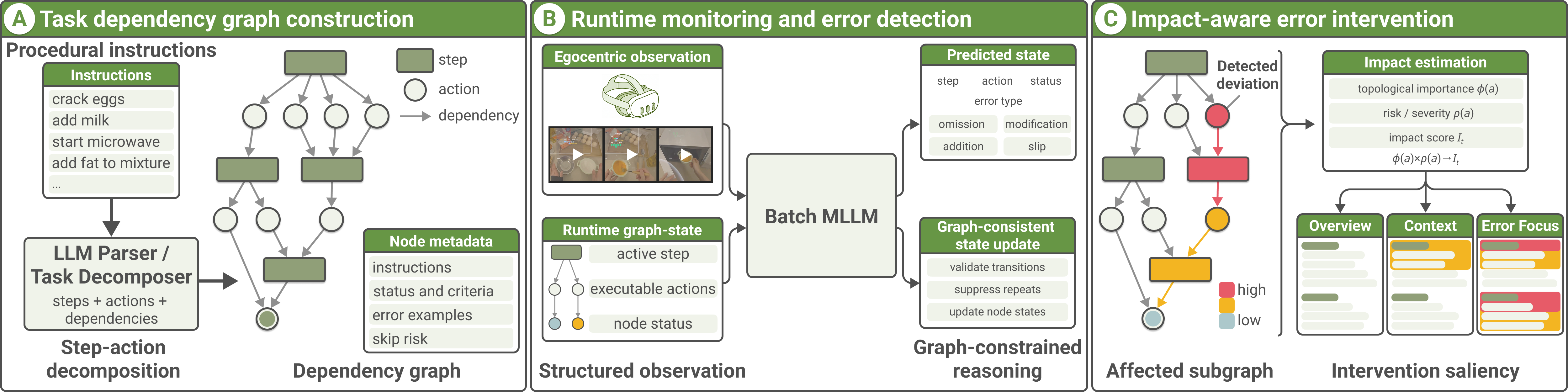}
    \Description{RegulAR System Overview. Figure 3A shows an MLLM parser converts instructions into a dependency-based DAG with action metadata. Figure 3B shows runtime monitoring utilizes egocentric video and a Batch MLLM to identify errors via a predefined taxonomy. Figure 3C shows impact-aware reasoning assesses node importance and downstream risks to determine intervention priority.}
    \caption{\textit{RegulAR} system overview. (A) An LLM parser converts procedural instructions into a hierarchical dependency graph and augments each action with status, error examples, and skip-risk metadata. (B) During execution, a batch MLLM reasons over recent egocentric frames and the current runtime graph state to predict the current step, action, and status. Graph-consistent state updates maintain task progress, infer omission from unmet prerequisites, and identify execution errors such as modification, addition, and slip. (C) The affected subgraph and associated risk are combined into an impact score that determines intervention saliency and drives overview, context, and error-focused assistance.}
    \label{fig:system}
\end{figure*}

\textit{RegulAR} is an AR assistant that converts multimodal procedural instructions into a structured task model and uses that model to interpret egocentric observations during execution. 
The system operates in two stages. 
First, an LLM parses instructions into a hierarchical dependency graph and augments each action with perceptual cues, representative error examples, and risk metadata (\textbf{DR1}; \Cref{fig:system}{A}). 
Second, a batch MLLM reasons over recent headset-captured frames and the current graph state at runtime to continuously monitor progress and \rw{automatically} detect errors. 
These deviations are classified by error types and estimated on their downstream impact (\textbf{DR2}; \Cref{fig:system}{B}).
Layered guidance is delivered through an adaptive HUD and voice interaction that scales intervention saliency to error significance (\textbf{DR3}; \Cref{fig:system}{C}). 
A lightweight live model supports low-latency spoken interaction throughout. 
This design allows \textit{RegulAR} to move beyond linear next-step prompting toward proactive error recognition and recovery support.

\subsection{Task Dependency Graph Construction}
\rw{To externalize task structure and state (\textbf{DR1}), we represent each task as a hierarchical directed acyclic graph (DAG). Unlike linear instruction lists, the graph distinguishes valid reordering from unmet prerequisites by separating semantic subgoals, observable actions, and their dependencies.}

Following Truong et al.~\cite{truong_twolevel_2021}, we decompose a procedure into high-level \textit{steps} and atomic \textit{actions}, making it easier for users to follow \textbf{(DR3)}. 
A step denotes a semantic subgoal (e.g., \textit{prepare the batter}), while an action corresponds to the smallest observable manipulation that can be monitored from egocentric video (e.g., \textit{add 3 tablespoons of flour into the mug}).  

Unless the instructions specify an explicit intra-step dependency, actions within the same step are treated as order-independent.

Inspired by G2Vid~\cite{nikita_g2vid_2022}, which presents all topological sorts of task execution order in one graph to weakly supervise error localization in videos, we generate task graphs based on step dependencies via an LLM parser to represent task structures \textbf{(DR1)}.
Formally, a task is represented as $G=(V,E)$ with two node types:
\[
V = V_S \cup V_A,
\]
where $V_S=\{S_1,\dots,S_N,S_{\text{done}}\}$ are step nodes and $V_A=\{A_{ij}\}$ are action nodes. Each step $S_i$ contains an action set $\mathcal{A}_i=\{A_{i1},\dots,A_{ik}\}$.

We use membership edges $E_M=\{(S_i,A_{ij}) \mid A_{ij}\in\mathcal{A}_i\}$ and represent step-level prerequisites with a predecessor set $\mathrm{Pred}(S_i)\subseteq V_S$, where $\mathrm{Pred}(S_i)$ contains the steps that must be completed before $S_i$ becomes executable. 
The executable action set at time $t$ is
\[
\mathcal{E}_t = \{A_{ij}\in\mathcal{A}_i \mid \mathrm{Pred}(S_i)\subseteq C_t^S,\; A_{ij}\notin C_t^A\},
\]
where $C_t^S$ and $C_t^A$ denote completed steps and completed actions. 
We construct the edge set $E = (S_i, A_{ij}) \cup (A_{ij}, S_k)$, such that each \textit{step} node $S_i$ connects to its action nodes $A_{ij}$, and each action node connects to subsequent steps whose dependency conditions become satisfied after completing $S_i$ (detailed prompt in Appendix~\ref{appendix:prompt1}). 
This formulation captures dependency constraints while allowing flexible execution order within and across independent task branches.

After the generation of graph structure, the LLM Parser generates perceptual criteria describing observable \textit{in-progress} and \textit{completion} conditions for each action node, inspired by Vid2Coach \cite{huh_vid2coach_2025} to improve visual grounding. 
We additionally generate representative examples for each error type and estimate skip-risk levels~\cite{lee_gtg_2025,kim_llmerror_2023} to facilitate error recognition.

\subsection{Runtime Monitoring and Error Recognition}
Since the task dependency graph is not linear, we need to keep track of actions and their statuses simultaneously to provide guidance at users' current progress. The problem is then formulated as follows: Given a sequence of frames from an egocentric video streaming, our goal is to infer the step sequence $\hat{Y}=(\hat{y}_1,\dots,\hat{y}_T)$ and the corresponding status sequence $\hat{S}=(\hat{s}_1,\dots,\hat{s}_T)$, where $\hat{y}_t$ denotes the predicted action label and $\hat{s}_t$ denotes the predicted status type at frame $t$. 
Each action is assigned one of the following states \textbf{(DR2)}: \textit{not\_started}, \textit{in\_progress}, \textit{complete}, or \textit{error}. 
To support structured error recognition, we adopt a taxonomy consisting of omission and execution errors~\cite{lee_egoped_2024}. 
An \textit{omission} occurs when a prerequisite action is skipped, and a later action is attempted before dependency conditions are satisfied. 
Execution errors occur when an intended action is performed incorrectly, and they further distinguish three execution-error types: 
\textit{modification}, where the action is performed in an alternative manner (e.g., \textit{using a different tool or ingredient}); 
\textit{addition}, where an extra action not represented in the task graph is introduced; 
and \textit{slip}, where the intended action fails to achieve the expected outcome (e.g., \textit{pouring liquid into the wrong container}). 

At runtime, \textit{RegulAR} maintains a set of executable actions $\mathcal{E}_t$ derived from the task dependency graph. 
Initially, only leaf actions are executable. The dependent \textit{action} nodes must be completed before proceeding to the next step. 
The actions are unlocked at the step level, meaning an action $A_{ij}$ becomes executable when all prerequisite steps specified by the task graph of $S_{i}$ are satisfied. 
Predicted observations are validated against the runtime state to prevent inconsistent transitions. 
For example, an action marked as \textit{complete} cannot revert to \textit{in\_progress} unless an error occurs.

\textit{RegulAR} adopts a batch MLLM for progress tracking and error recognition~\cite{chang_worldscribe_2024}.
At runtime, \textit{RegulAR} performs monitoring in fixed decision cycles rather than frame-by-frame recognition. 
Every 5\,s, the system samples a short frame window from the headset camera at 1\, fps and combines it with a runtime snapshot of the task graph.
\rc{This decision interval follows prior work using periodic sampling and multimodal reasoning for task progress tracking~\cite{huh_vid2coach_2025}. As no widely adopted sampling standard exists for this setting, we adopt a 5\,s decision cycle as a practical trade-off between recognition accuracy, responsiveness, latency, and computational cost.}
\rw{The sampled frame window was intended to provide temporal context for continuous actions while maintaining interactive performance.}
At decision cycle $t$, the batch MLLM receives the recent frame window $W_t$ together with the current graph state, including the active step, executable actions $\mathcal{E}_t$, and node statuses.
It returns a structured observation
\[
z_t = (\hat{v}^S_t,\hat{v}^A_t, \hat{s}_t)
\]
where $\hat{v}^S_t \in V_S$ denotes the predicted step, $\hat{v}^A_t \in V_A$ denotes the predicted action, and $\hat{s}_t$ denotes the execution status. 
Execution errors (\textit{modification}, \textit{addition}, and \textit{slip}) are obtained directly from the MLLM output.
Omission errors are inferred from the graph state when the predicted action is not currently executable, i.e., $\hat{v}^A_t \notin \mathcal{E}_t$.
Predicted observations are then validated against the runtime graph before being committed. 
Repeated identical predictions are suppressed, completed actions are not reopened unless a later error invalidates them, and only graph-consistent transitions update node state. 
This graph-consistent update reduces oscillations in the unstable state during streaming inference.

\subsection{Impact-Aware Error Intervention}
Not all errors require the same level of intervention. 
\textit{RegulAR} therefore separates intervention into two decisions: \emph{what} to communicate and \emph{how strongly} to communicate it. 
Error type determines the content of guidance, while estimated downstream impact determines intervention saliency.
This echoes prior work on adaptive AR guidance~\cite{cho_evalxrmessagenotification_2025}, which suggests tailoring feedback strength and content to balance informativeness and minimal disruption \textbf{(DR3)}.

\textbf{\textit{Topological Importance and Impact Estimation.}}
To estimate how strongly an error may affect the overall procedure, the system evaluates the structural importance of the affected actions and the error risk score provided by MLLM. 
For an action node $a\in V_A$, we define structural importance $\phi(a)$ using downstream reachability and remaining task distance~\cite{kelley_path_1961, kelley_pathplan_1959, qiao_longpath_2025}:
\[
\phi(a)=
\alpha\frac{|\mathrm{desc}(a)|}{\max_b|\mathrm{desc}(b)|}
+
(1-\alpha)\frac{\mathrm{dist}(a,S_{done})}{\max_b\mathrm{dist}(b,S_{done})}
\]
where $\mathrm{desc}(a)$ is the set of downstream actions reachable from $a$, $\mathrm{dist}(a,S_{\text{done}})$ is the longest-path distance from $a$ to task completion, and $\alpha=0.5$ balances breadth of downstream influence against remaining task distance.
Intuitively, actions that occur earlier in the procedure or influence many downstream operations receive higher importance values.

When an error is detected at time $t$, the system determines the set of affected actions $U_t$. For omission errors, let $\mathrm{pred}(\hat{v}^A_t)$ denote the set of prerequisite actions that must be completed before the predicted action $\hat{v}^A_t$ becomes executable.
For execution errors, the affected set includes the current erroneous action and any completed actions whose complete status may be affected by this error, provided by MLLM:
\[
U_t=
\begin{cases}
\{a\in\mathrm{pred}(\hat{v}^A_t)\mid a \text{ is not completed}\}, & \text{if omission error},\\
\{\hat{v}^A_t\}\cup \mathrm{desc}(\hat{v}^A_t), & \text{if execution error}.
\end{cases}
\]
Each affected action $a\in U_t$ is assigned a risk value $\rho(a)$. 
For omission errors, $\rho(a)$ comes from the skip-risk value specified in the task graph; for execution errors, it is estimated using the subtype-specific risk weight together with the severity score predicted by the batch MLLM. We compute a local impact term
$r(a)=\phi(a)\rho(a)$
and aggregate these risks as
\[
I_t = 1-\prod_{a\in U_t}(1-r(a))
\]
This bounded aggregation allows multiple small errors to accumulate while preventing the impact score from growing unbounded with the number of affected nodes.

\textbf{\textit{Interface Adaptation.}}
To support different stages of task execution, \textit{RegulAR} provides three graph views following the principle of overview first and context-dependent detail.
The overview mode presents the full node-link graph to help users understand the task structure. 
The context mode provides a glanceable task-state overlay highlighting the current step, completed actions, and recommended next actions \rc{(\Cref{fig:ui_comparison}{b})}.
When errors occur, the interface switches to an error mode that highlights affected actions and recovery paths \textbf{(DR2)}, helping users quickly re-establish task context. 
The interface adapts visual saliency based on estimated impact. 
During normal execution, the overlay remains transparent to minimize distraction, while higher-impact situations trigger more visually salient presentations. 
Inspired by prior work that adjusts interface transparency based on risk levels \cite{pei_attentionar_2025}, we map impact levels to three discrete transparency settings rather than continuous adaptation to avoid flicker and preserve readability during execution.

\textbf{\textit{Error-Aware Assistance.}}
Rather than applying uniform feedback, \textit{RegulAR} tailors assistance based on error type to support effective recovery \textbf{(DR2)}. 
Assistance messages are structured around three elements (what happened, why it matters, and how to recover) across four error types~\cite{lee_egoped_2024}.
For \textit{omission}, the system highlights missing prerequisite actions and recommends returning to the earliest required step to restore task consistency. 
For \textit{slip}, the system identifies incorrect execution and suggests correcting the current step before proceeding. 
For \textit{addition}, the system evaluates whether the extra action affects task outcomes and intervenes only when the estimated impact is non-negligible. 
For \textit{modification}, the system assesses whether alternative procedures affect correctness and recommends reverting to the prescribed method when necessary.

\textit{RegulAR} continuously maintains a recommended next action based on the current task graph and runtime state \textbf{(DR1)}. 
To preserve procedural continuity in partially ordered tasks, the system favors local continuity over globally re-ranking all executable actions.
It first preserves the current \textit{in-progress} action when possible, then prioritizes recovery from execution errors, then selects the earliest unfinished action in the current executable step, and finally advances to the next executable step if the current one is complete. 
If no forward candidate is available, the system falls back to the earliest deferred prerequisite action. 
This policy maintains local procedural consistency while respecting dependency constraints in the task graph.
When task states change, \textit{RegulAR} updates the recommended action and provides proactive guidance. 
Assistance messages are structured around three elements: \emph{what happened}, \emph{why it matters}, and \emph{how to recover}. 
Users may also issue queries at any time. 
A lightweight live model is adopted to answer them verbally using the runtime task graph together with the recent egocentric frame window as contextual grounding.

\subsection{Implementation}
We implemented the real-time assistant as a Unity 2022.3.60f1 application running on Meta Quest 3, with a serverless backend hosted on Modal~\cite{modal} for model serving, communication, and data routing. 
The LLM Parser in task dependency graph construction employs GPT-5.
The system integrates with the Google Gemini Multimodal Live API \cite{google_multimodal_live}, which supports low-latency bidirectional voice and video streaming over WebSocket. 
We used gemini-2.5-flash-native-audio-preview-12-2025 for real-time spoken interaction and gemini-2.0-flash-lite for batch inference. 
Egocentric video and interaction data captured on the headset are streamed to the backend for real-time progress monitoring and feedback generation.
\section{User Evaluation}
To evaluate the effectiveness of \textit{RegulAR} in supporting error recognition and procedural task assistance, we conducted a within-subject user study comparing our graph-grounded guidance with a baseline MLLM-based task guidance system without \rc{explicit} dependency graph regulation.
We examined whether \textit{RegulAR} would (1) improve \rw{users’ procedural understanding and system state tracking}, (2) provide effective support for error recovery, and (3) maintain comparable or lower perceived task demand during execution.

\subsection{Method}
\textbf{Participants.}
We recruited 12 participants (P7-P18, 6 female, 6 male; age: $M = 24.08$, $SD = 2.27$) from the university community through social media, reporting moderate familiarity with AR technology ($M = 2.25$) on a 5-point Likert scale. 

\begin{figure}[t]
  \centering
  \includegraphics[width=\linewidth]{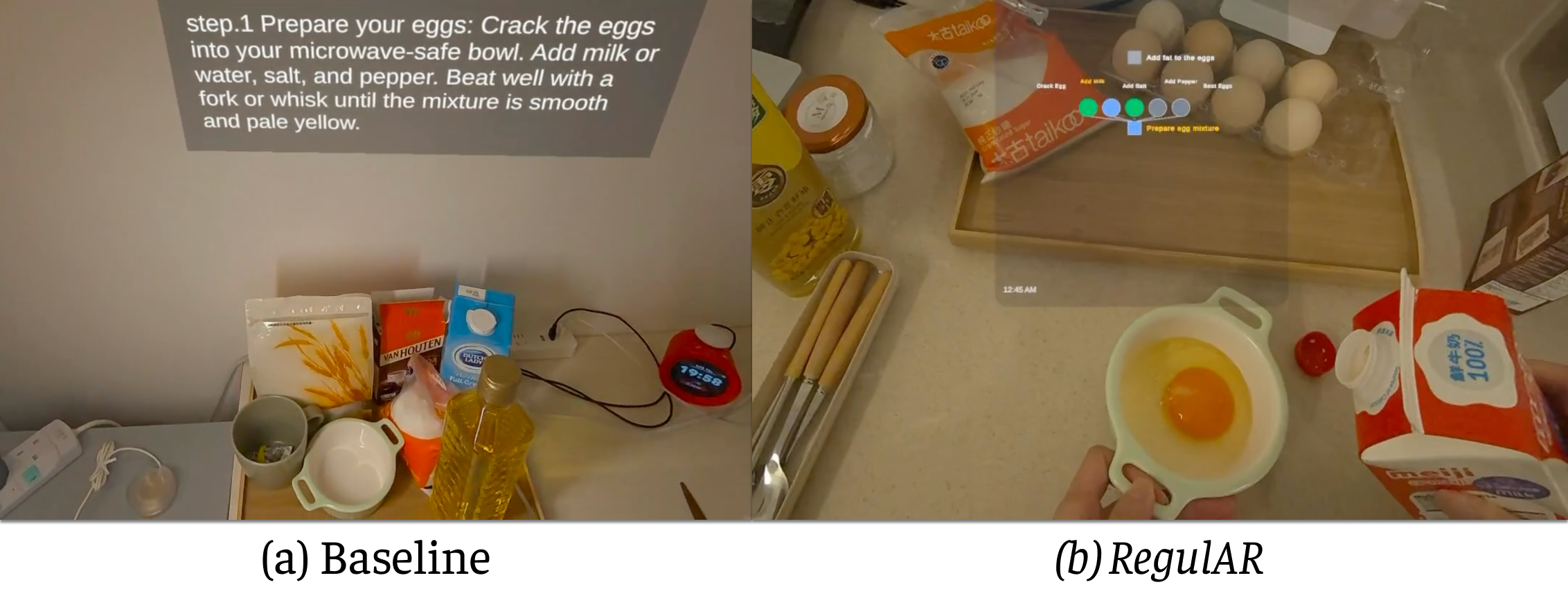}
  \caption{\rc{Actual headset screenshots of (a) the baseline displaying current-step guidance and (b) \textit{RegulAR}'s context mode visualizing task progress, completed actions, and recommended next actions.}}
  \Description{A two-panel comparison of actual headset interfaces used in the study. The baseline view presents only the current procedural step as textual guidance over the physical workspace. The RegulAR view shows a task dependency graph with completed, current, and available actions, together with the recommended next action.}
  \label{fig:ui_comparison}
\end{figure}

\rw{
\textbf{Baseline.}}
Each participant completed one task with \textit{RegulAR} and one with a baseline MLLM-based guidance system. Both conditions used the same \rc{procedural documents}, MLLM, egocentric visual input, monitoring interval, and \rc{AR headset}.
\rc{The procedural documents were imported from the same instruction source (illustrated by the tablet in \Cref{fig:teaser}). The primary difference was the use of an explicit task graph. \textit{RegulAR} maintained a task graph for reasoning, state tracking, and visualization, whereas the baseline relied solely on prompt-based reasoning over textual instructions without an explicit task model.}
Consequently, \rw{the baseline displayed only current-step information and generated guidance directly from the instructions.}
\rc{Both conditions were implemented in AR to provide hands-free, real-time assistance during physical task execution. \Cref{fig:ui_comparison} shows the actual interfaces used in the two conditions.}

\textbf{Procedure.}
The study\footnote{The protocol was approved by the IRB at our institution} was conducted in a campus kitchen equipped with microwaves and induction cookers. After consent and a 5-minute tutorial for familiarization, each participant completed two tasks, Microwave Scrambled Eggs~\cite{pathculture_microwave_eggs_2025} and Scented Candle Making~\cite{chapman_candles_2024}, one per condition, with task assignment and condition order counterbalanced. Both tasks involved multi-step execution, physical object manipulation, and structured step dependencies, while differing in structure (linear vs. partially dependent). Participants reported low prior familiarity ($M=3.08$ and $M=2.58$, respectively) on a 7-point Likert Scale, increasing the likelihood of deviations. To approximate real-world uncertainty, we introduced imperfect material availability (e.g., missing items or misleading alternatives such as cocoa powder instead of pepper). Each task was designed to be completed within 15 minutes while still allowing for realistic breakdowns. We further conducted semi-structured interviews to probe participants’ experiences and perceptions of system support for error recognition and recovery. The total session duration for each participant was approximately one hour.

\textbf{Measures and Analysis.}
After each task, participants completed \rw{the six NASA-TLX subscales~\cite{hancock_nasa_1988}, adapted to 7-point Likert scales, along with} 7-point ratings covering task understanding, progress awareness, error awareness, explanation quality, recovery support, confidence, and overall preference (\Cref{tab:questions} lists the questionnaire items).
\rc{We analyzed the NASA-TLX subscales separately and did not apply the original pairwise weighting procedure.}
For the Likert-scale responses, we report descriptive statistics (mean and standard deviation).
Given the ordinal and paired nature of the data, we used Wilcoxon signed-rank tests to assess differences between the two conditions.
\rc{All statistically significant findings remained significant after Benjamini--Hochberg FDR correction~\cite{benjamini_1995}.}
The semi-structured interview \rw{audio recordings} were transcribed and analyzed using thematic analysis~\cite{Braun01012006Thematic}, focusing on task understanding, error interpretation, and responses to system interventions.
In addition, headset recordings and system logs were reviewed to identify task breakdowns, error-recognition accuracy, and recovery behaviors, providing complementary behavioral evidence to the subjective reports.
\rc{Given the small sample size and multiple comparisons, we treat the quantitative findings as exploratory and interpret them together with qualitative and behavioral evidence.}

\subsection{Findings}
\begin{figure}[t]
  \centering
  \includegraphics[width=\linewidth]{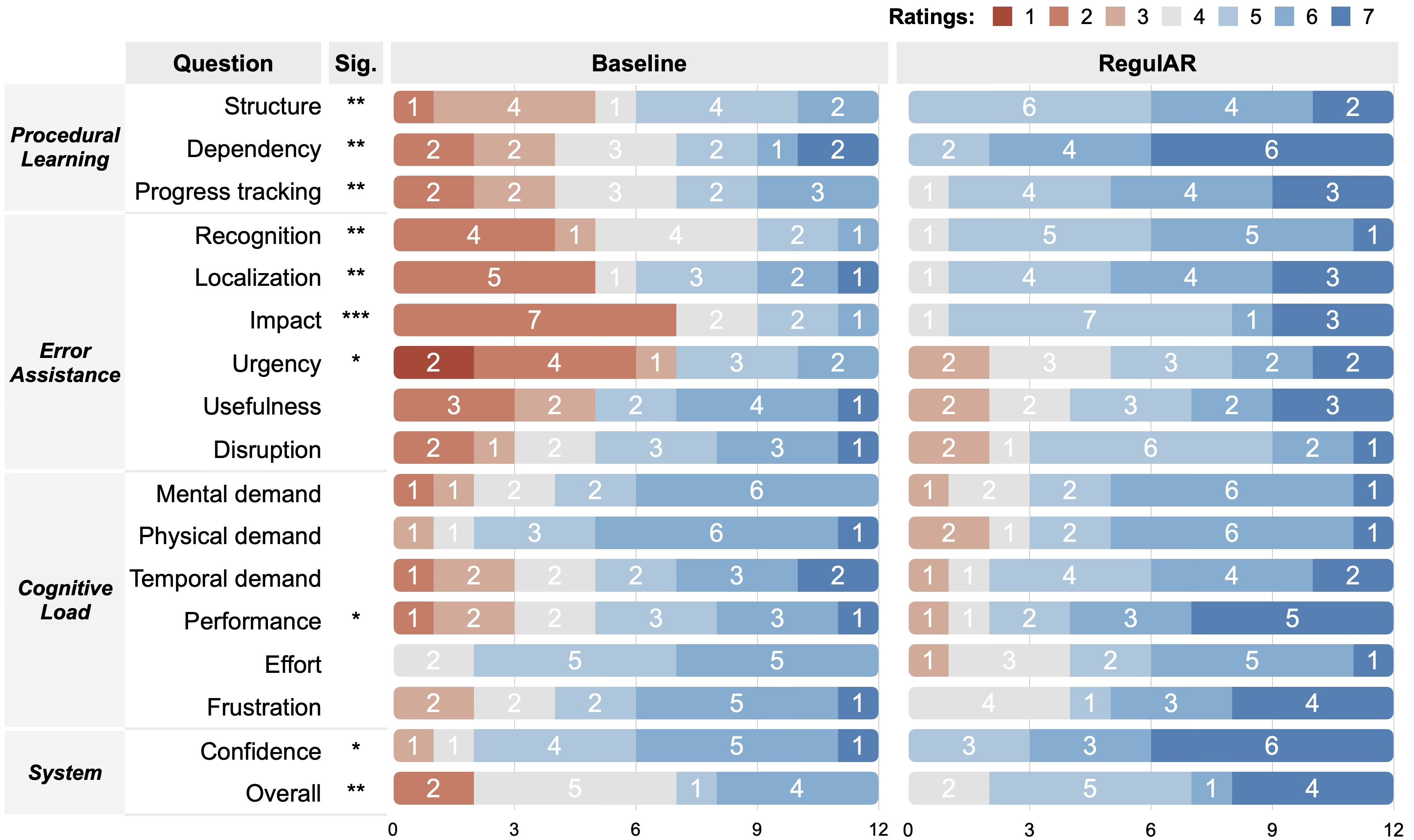}
  \caption{Distribution of questionnaire results for baseline and \textit{RegulAR} (1 = negative, 7 = positive). Asterisks indicate statistical significance based on Wilcoxon signed-rank tests for paired samples: * $p < .05$, ** $p < .01$, *** $p < .001$.}
  \Description{Distribution of questionnaire ratings for Baseline and RegulAR across procedural learning, error assistance, cognitive load, and system measures, based on Wilcoxon signed-rank tests for paired samples. Ratings range from 1 to 7, where higher values indicate more positive responses. RegulAR shows higher ratings for procedural learning, including structure, dependency, and progress tracking, and for error assistance, including recognition, localization, impact, urgency, usefulness, and disruption. Cognitive load measures, including mental demand, physical demand, temporal demand, effort, and frustration, are similar between conditions, while RegulAR shows higher perceived performance. System-level ratings show higher confidence and overall ratings for RegulAR.}
  \label{fig:user_eval_survey}
\end{figure}
\rw{Overall, participants reported better understanding of task structure, greater progress awareness, and better support for error recovery when using \textit{RegulAR}, without a measurable increase in workload (\Cref{fig:user_eval_survey}). 
These patterns were accompanied by higher confidence and an overall preference for \textit{RegulAR}.}

\subsubsection{Task Structure and Progress Awareness}
~

\textbf{Finding 1: \rw{Participants reported a better understanding of task structure when using the dependency graph.}}
Participants reported significantly better understanding of overall task structure ($M=6.00$, $SD=1.56$ vs. $M=4.25$, $SD=1.60$; $Z=2.67$; $p<.01$) and step dependencies ($M=6.17$,$SD=1.72$ vs. $M=4.17$, $SD=1.27$; $Z=2.67$; $p<.01$) when using \textit{RegulAR}. Interview data suggest that the graph provided participants with more than a display of steps---it offered a shared view of the task structure that helped them anticipate upcoming actions and understand how individual steps related to the overall goal. Participants described a consistent pattern: orienting themselves via the graph, focusing on the physical task while listening to voice guidance, and returning to the graph when the task state changed. As P15 commented, \textit{``The graph serves naturally like navigation, and completing steps brings a sense of achievement, like playing a game.''} In contrast, the baseline struggled to maintain an accurate representation of task state, leaving participants to rely on sequential text and memory alone, \textit{``it didn't know what I was doing''} (P8, P12, P15, P17).

\textbf{Finding 2: Explicit state tracking reduced disorientation and prevented structurally inconsistent guidance.}
Participants rated progress tracking significantly higher with \textit{RegulAR} ($M=6.17$, $SD=1.72$ vs. $M=4.17$, $SD=1.27$;
$Z=2.67$; $p<.01$). By continuously encoding completed actions into the graph state, \textit{RegulAR} kept users synchronized with the system's evolving understanding of the task. Without this, the baseline frequently lost track of prior actions. P12 relied solely on text instructions due to limited trust in voice guidance, while P14 repeatedly received suggestions to add milk even though it had already been added, eventually over-diluting the egg mixture. These cases \rw{suggest that} the absence of explicit state tracking may contribute to structurally misaligned guidance and compound errors over time.

\subsubsection{Error Interpretation and Recovery}
~

\textbf{Finding 1: \rw{Participants reported better error recognition and localization with graph-grounded guidance.}}
\rw{Participants reported significantly higher ratings for error recognition} ($M=6.08$, $SD=1.38$ vs. $M=4.00$, $SD=1.65$; $Z=2.63$; $p<.01$) and \rw{localization} ($M=5.92$, $SD=1.91$ vs. $M=4.08$, $SD=1.68$; $Z=2.59$; $p<.01$) \rw{with \textit{RegulAR}}, improvements also reflected in objective recognition performance (\Cref{sec:error_accuracy}).
Participants attributed these gains to the error mode, which highlighted the affected portion of the task graph when an error occurred, allowing participants to directly see which steps were involved and where recovery was needed to begin (P10). By contrast, the baseline indicated that something was wrong without anchoring the issue to a specific point in the task structure, leaving participants to reconstruct the problem context themselves.

\textbf{Finding 2: Structured guidance helped users reason about downstream consequences and prioritize recovery.}
\rw{Participants also reported better support for understanding downstream consequences} ($M=6.25$, $SD=1.53$ vs. $M=3.92$, $SD=1.56$; $Z=2.93$; $p<.001$) and \rw{assessing} their urgency ($M=5.83$, $SD=1.92$ vs. $M=4.17$, $SD=1.40$; $Z=2.51$; $p<.05$). By organizing feedback around what happened, why it mattered, and how to recover, \textit{RegulAR} gave participants a basis for reasoning about recovery rather than simply reacting to isolated instructions. As P8 summarized, \textit{``Graphs provide glanceable, timely, and minimally intrusive assistance\rw{.''}} In contrast, the baseline focused on suggesting the next action without explaining the underlying error or its implications, leaving P16 uncertain and less confident when mistakes occurred.

\subsubsection{Guidance Informativeness and Disruption}
~

\textbf{Finding 1: \textit{RegulAR} maintained comparable cognitive load while \rw{participants reported better task performance.}}
NASA-TLX scores showed no significant differences between conditions on mental demand ($M=5.67$, $SD=1.21$ vs. $M=5.33$, $SD=1.15$; $Z=0.85$; $p=0.52$), physical demand ($M=5.83$, $SD=1.08$ vs. $M=5.25$, $SD=1.29$; $Z=0.59$; $p=0.77$), temporal demand ($M=5.83$, $SD=1.64$ vs. $M=5.42$, $SD=1.16$; $Z=1.12$; $p=0.31$), and effort ($M=5.25$, $SD=0.75$ vs. $M=5.17$, $SD=1.19$; $Z=0.18$; $p=0.98$), indicating that the graph representation did not impose a measurable workload penalty. Nonetheless, participants rated their task performance significantly higher with \textit{RegulAR} ($M=5.67$, $SD=1.50$ vs. $M=4.67$, $SD=1.15$; $Z=2.37$; $p<.05$). P11 noted an initial learning cost in understanding the graph structure, but expected this cost to diminish with familiarity and considered structured guidance more beneficial for complex tasks.

\textbf{Finding 2: Impact-based filtering improved guidance quality, but a tension between proactivity and autonomy persisted.}
Although perceived guidance usefulness ($M=5.75$, $SD=1.88$ vs. $M=5.08$, $SD=1.44$; $Z=1.42$; $p=0.18$) and disruption ($M=5.92$, $SD=1.62$ vs. $M=5.67$, $SD=1.37$; $Z=0.58$; $p=0.58$) did not differ significantly, qualitative accounts revealed meaningful differences in guidance quality. The baseline exhibited inconsistent behavior, sometimes missing consequential deviations and at other times issuing repeated notifications regardless of severity; P8 and P12 found rapid, successive suggestions distracting. 
\textit{RegulAR}'s impact-based filtering reduced such noise, \rc{although six participants} \rw{noted recognition latency, with some interventions arriving after the relevant moment had passed.}
Sources of frustration also differed across conditions: baseline frustration stemmed from unreliable guidance that undermined confidence (P12), while \textit{RegulAR} frustration was tied to latency disrupting task flow. Beyond these system-level issues, participants revealed divergent preferences: some preferred autonomy and on-demand assistance (P12), while others valued continuous, proactive guidance (P7, P8), highlighting the need for adaptive, user-controllable guidance strategies.

\subsubsection{User Experience and Preference}
~

\textbf{Finding 1: \textit{RegulAR} reduced anxiety about making mistakes and increased execution confidence.}
Participants reported significantly higher confidence when using \textit{RegulAR} ($M=5.33$, $SD=1.07$ vs. $M=4.25$, $SD=1.42$; $Z=2.34$;
$p<.05$), describing feeling \textit{``more assured''} and \textit{``less worried about making mistakes''} (P12, P13), particularly in unfamiliar tasks where the perceived cost of errors was higher. This increased confidence was closely tied to \textit{RegulAR}'s ability to maintain a clear representation of task state; knowing where they were and what had been completed reduced the uncertainty that typically accompanies error-prone execution.

\textbf{Finding 2: Participants \rw{generally} preferred \textit{RegulAR} and anticipated greater benefits in more complex tasks.}
Participants expressed a significantly stronger overall preference for \textit{RegulAR} ($M=5.42$, $SD=1.44$ vs. $M=4.25$, $SD=1.22$; $Z=2.53$; $p<.01$), with 10 of 12 indicating they would choose \textit{RegulAR} for future procedural tasks. Preference was especially pronounced for complex scenarios where structured, error-aware support would be most beneficial. P17 noted that \textit{RegulAR} compared favorably to video or mobile-based instructions that require manual scrolling, which is impractical when hands are occupied, underscoring the value of in-situ, state-aware guidance during hands-on execution.
\section{Technical Evaluation}
To evaluate the robustness of our proposed system, we analyzed three key requirements for an egocentric procedural AI assistant from \cite{li_egoproce_2026}: \rw{procedural learning, error recognition}, and the quality of error assistance.
\rc{We first evaluate procedural learning through the quality of the generated task graphs, followed by runtime evaluations of procedural learning, error recognition, and error assistance.}

\subsection{\rc{Graph Generation}}
\rc{\textbf{Method.}
We tested the graph generation pipeline on 10 procedural documents spanning recipes, drink preparation, craft/science procedures, and meeting setup (\Cref{tab:instructions}). We first automatically checked structural validity, including unique node IDs, acyclicity, and valid dependencies. We then manually audited semantic quality across four levels: hallucinated content, step-segmentation errors, action-node errors, and dependency-edge errors. We also evaluated whether the generated criteria were visually grounded and matched the corresponding action.
}

\rc{
\textbf{Results.}
All 10 generated graphs were structurally valid DAGs, covering 118 steps and 176 actions. 
In the manual audit, hallucination was low ($M=0.7\%$, $SD=1.5\%$), while step segmentation, action-node, and dependency-edge errors remained limited ($M=2.1\%$, $SD=3.7\%$; $M=5.2\%$, $SD=3.8\%$; and $M=3.3\%$, $SD=4.5\%$, respectively). Completion criteria were correctly captured in most cases ($M=92.2\%$, $SD=3.9\%$). Most remaining issues came from compact representations of repeated, optional, or concurrent actions.
}

\subsection{Procedural Learning \& Error Recognition}
\label{sec:error_accuracy}
\begin{table}[t]
\centering
\small
\setlength{\tabcolsep}{3pt}
\begin{tabular}{@{}lcccccccc@{}}
\toprule
& \multicolumn{4}{c}{\textit{RegulAR}}
& \multicolumn{4}{c}{Baseline} \\
Category
& \rw{Acc./Recall}
& F1
& \rc{FP}
& \rw{GT}
& \rw{Acc./Recall}
& F1
& \rc{FP}
& \rw{GT} \\
\midrule
Action
& \textbf{0.90}
& --
& --
& 535
& 0.69
& --
& --
& 525 \\
\midrule
In-progress
& \textbf{0.90}
& \textbf{0.93}
& 15
& 327
& 0.86
& 0.88
& 33
& 327 \\
Complete
& \textbf{0.90}
& \textbf{0.86}
& 18
& 93
& 0.46
& 0.61
& 2
& 50 \\
\midrule
Omission
& \textbf{0.76}
& \textbf{0.76}
& 5
& 21
& 0.67
& 0.32
& 45
& 18 \\
Slip
& \textbf{0.71}
& \textbf{0.77}
& 1
& 7
& 0.50
& 0.50
& 4
& 8 \\
Modification
& \textbf{0.80}
& \textbf{0.81}
& 10
& 59
& 0.67
& 0.62
& 15
& 30 \\
Addition
& 0.91
& \textbf{0.77}
& 10
& 22
& \textbf{1.00}
& 0.65
& 17
& 16 \\
\bottomrule
\end{tabular}

\caption{
\rc{Action accuracy (Acc.)}, \rw{recall}, F1 score, \rc{false-positive (FP) counts}, and \rw{ground-truth instances (GT)} for procedural state tracking and error recognition. \rc{The Acc./Rec.\ column reports accuracy for Action and recall for all other categories. FP reports false-positive detections (hallucinations). Boldface indicates better performance.}
}
\Description{
Comparison of RegulAR and the baseline on action recognition, procedural state tracking, and error recognition. The Accuracy-or-Recall column reports action accuracy for the Action row and recall for all other categories. F1 score, false-positive counts, and ground-truth instances are reported where applicable. False positives represent hallucinated recognitions. RegulAR performs better for most categories, while the baseline achieves higher recall for addition recognition.
}
\label{tab:recognition}
\end{table}
\begin{figure}
    \centering
    \includegraphics[width=\linewidth]{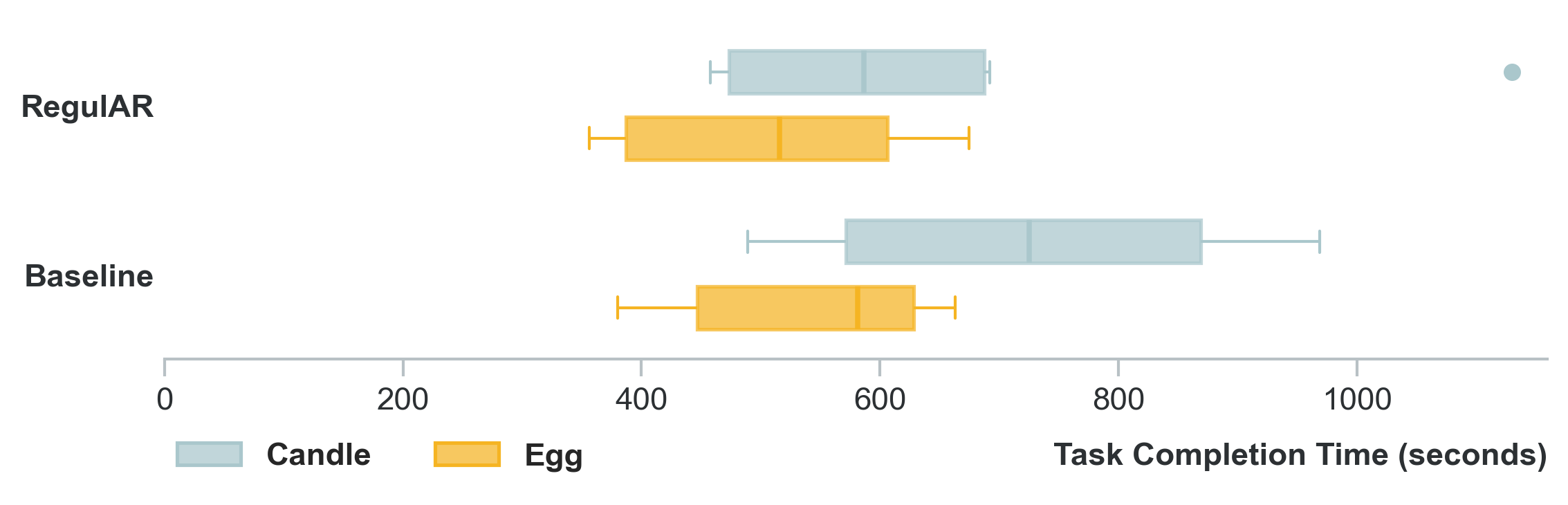}
    \caption{Task completion time across Baseline and RegulAR conditions for the candle task and the egg task. \rc{The circle on the right indicates an outlier.}}
    \label{fig:completion_time}
    \Description{Boxplots of task completion times in seconds for Baseline and RegulAR across two procedural tasks, egg and candle making. Completion times are generally longer for the candle task than the egg task in both conditions. Compared with the baseline, median completion times decrease from 581.5 to 516 seconds for the egg task and from 725 to 586.5 seconds for the candle task under RegulAR, with a larger reduction for the candle task; differences are not statistically significant. The RegulAR candle condition shows a high outlier, while the other conditions exhibit relatively comparable spreads.}
\end{figure}
\textbf{Method.}
We collected and evaluated logs \rw{from 12 user study sessions (24 task executions) involving} action recognition, step tracking, and error recognition across multiple procedural tasks. 
\rw{We report accuracy for action recognition and recall, F1 score, and false-positive (FP) counts for procedural states and error types.}
\rc{Recall measures the proportion of correctly recognized ground-truth instances, FP reports false-positive detections (hallucinations), and F1 summarizes the balance between missed detections and hallucinations.}
To avoid evaluation bias for durative actions (e.g., \textit{“microwave the egg mixture on high for 20 seconds”}) that span across multiple time intervals and thus \rw{receive} multiple predictions, we limited the evaluation to at most two consecutive frames. \rw{Repeated observations were merged during evaluation, resulting in fewer effective ground-truth instances.}
The sum of status and error types may not equal the number of actions, because \rw{status and error correctness was not evaluated when the corresponding action was recognized incorrectly.}

\textbf{Results.}  
We evaluated \rw{535 and 525 ground-truth observations for \textit{RegulAR} and the baseline, respectively (\Cref{tab:recognition}).} Although \textit{RegulAR} resulted in slightly shorter completion times (\Cref{fig:completion_time}), more observations were sampled because baseline interactions often remained in the same state for longer periods.
\rc{The single outlier was associated with network delay.}
Among error types, modifications were the most common, while slips were the least.

Overall, graph-constrained reasoning improved both procedural \rw{learning} and error recognition compared to textual prompting alone.
In terms of \rw{recall}, the improvement mainly comes from better contextual grounding: the dependency graph constrains predictions to contextually valid actions and expected transitions, reducing regressions to previously completed steps and improving completion recognition.
From the F1 perspective, the largest gain appears in omission recognition. The baseline lacks task-dependency reasoning and often misclassifies harmless reordering or intermediate actions as omissions (e.g., asking the user to melt the wax first, even if securing the wick is executable), leading to a lower F1 score. 
\rc{This behavior is also reflected in the FP counts: without explicit task dependencies, the baseline frequently hallucinates omission errors, whereas \textit{RegulAR} substantially reduces these false positives through graph-grounded reasoning.}
In contrast, \textit{RegulAR} distinguishes unmet prerequisites from valid alternative orderings, resulting in more balanced precision and recall.
Execution-level errors showed moderate improvements. 
Both systems occasionally misclassified normal actions as additions, especially when the action was not explicitly represented in the graph (e.g., operating the cooktop being treated as an extra step). 
However, \textit{RegulAR} maintained a more balanced overall detection performance, reflected in higher F1 scores \rc{and consistently lower FP counts across most error categories.}

\subsection{Error Assistance}
\begin{figure*}[t]
    \centering
    \includegraphics[width=\textwidth]{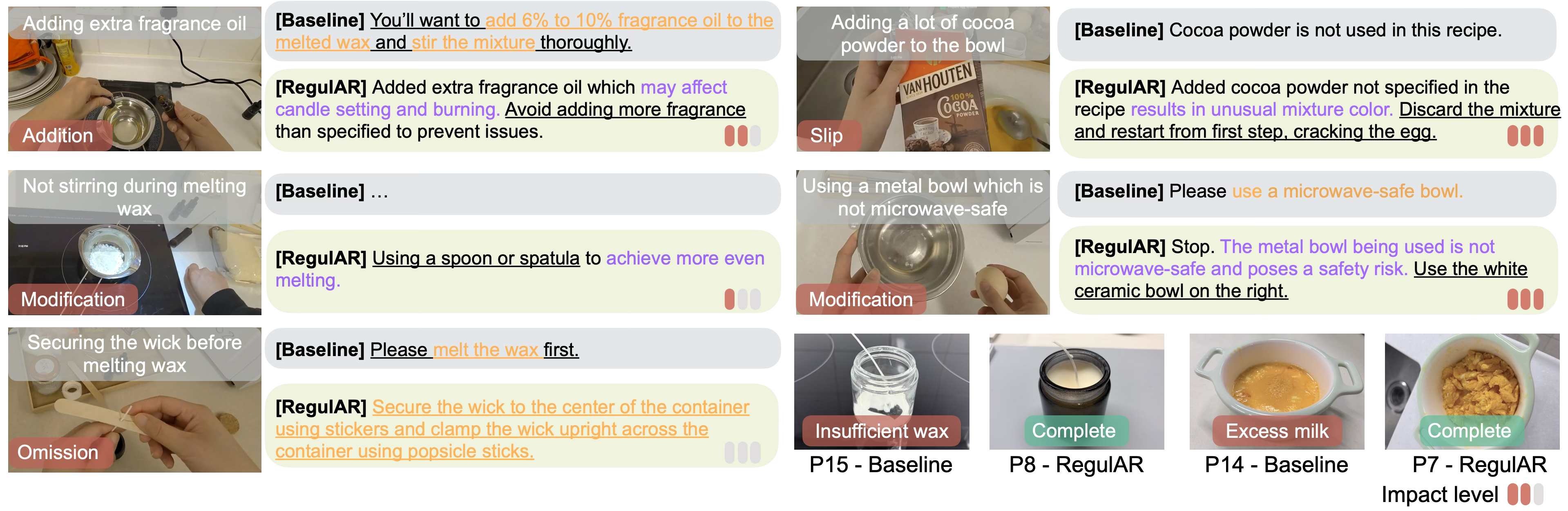}
    \caption{Qualitative examples and final \rw{outcomes} of \textit{RegulAR} and Baseline during the user study. Baseline tended to repeat the instruction (orange) and briefly outlined the next step (underline), while \textit{RegulAR} explained the reasons and risks (purple).}
    \Description{Comparison of Baseline and RegulAR assistance across procedural errors in the candle and microwave scrambled egg tasks, including addition, slip, modification, and omission. Each example shows user actions with corresponding feedback from both systems. RegulAR provides more specific explanations, recovery guidance, and impact levels, while Baseline responses are brief and sometimes contain recognition errors. The bottom row shows task outcomes across participants: unsuccessful outcomes, such as insufficient wax and excess milk, occur under Baseline, while successful outcomes are achieved with RegulAR.}
    \label{fig:qualitative_examples}
\end{figure*}
\Cref{fig:qualitative_examples} presents representative examples of error assistance from the baseline and \textit{RegulAR} across different error types collected during the user study.
These examples illustrate how \textit{RegulAR} follows task dependencies rather than predefined step order and adapts guidance based on the impact level of the detected error.
Baseline guidance tended to repeat instruction snippets and provided brief feedback focusing only on what had occurred or what to do next.
\textit{RegulAR} provided more structured assistance that explained the error, why it needed to be addressed, and how to recover, with varying length and urgency of tone based on impact level (e.g., stopping the user before the explanation).
In a case where the user used a spoon instead of a fork to stir the egg mixture, the baseline did not provide feedback. \textit{RegulAR} instead applied a low-impact intervention, suggesting that using a fork or whisk would improve aeration. This demonstrates \textit{RegulAR}’s ability to provide optional, low-urgency assistance for minor errors.
For slip errors requiring recovery, \textit{RegulAR} further identified the appropriate rollback step. When a user added cocoa powder, \textit{RegulAR} explained the error, described its impact on the mixture, and instructed the user to restart from the egg-cracking step. 
In contrast, the baseline only indicated that an error had occurred without suggesting how to recover.
\section{Discussion}
Drawing on the \rw{needs-finding} study and evaluation results, we derive five takeaways for proactive, error-aware AR task assistants. 
\subsection{Design Implications}

\textbf{AR guidance should be designed around recoverability rather than next-step compliance alone.}
A central implication of our findings is that the critical moments in procedural work are not only when users need the next instruction, but also when they need to know whether the current state is still valid and how to proceed after a deviation. 
In the \rw{needs-finding} study, participants repeatedly asked questions such as "\textit{Is this right?}" and "\textit{Can I move on?}", indicating that procedural support is as much about validating state and maintaining recoverability as it is about delivering the next step. 
This suggests a broader design target for AR guidance: in addition to \rw{helping users stay on an ideal path}, help them remain able to reach a valid end state after breakdowns. 
Similar to navigation systems that redirect users after wrong turns~\cite{amores_navigation_2021,jimenez_navigation_2019}, procedural AR assistants should be designed to support recovery before and after errors occur, not only to prevent them.

\textbf{Procedural errors are relational and must be interpreted in context.}
Our results show that procedural errors are difficult to understand from local action recognition alone. 
Their meaning depends on unmet prerequisites, valid state transitions, and how the deviation affects later progress. 
This interpretation is supported by the technical evaluation, which shows that \textit{RegulAR} achieved its greatest improvements in completion and omission decisions when explicit task dependencies matter more than \rw{the appearance of actions.}
The user study complements this pattern: participants using \textit{RegulAR} reported better understanding of where errors occurred, what they affected, and how urgent they were. 
These findings suggest that error support should not treat all deviations as equivalent. 
Instead, it should distinguish among different error types and communicate them in terms of consequence and recoverability, helping users understand a deviation before prescribing a fix.

\rc{\textbf{A visible task model can serve as a shared state contract between the user and the assistant.}
The dependency graph supported two related functions in \textit{RegulAR}. 
Computationally, it constrained model predictions to procedurally plausible states and transitions. 
In the interface, it externalized the same state representation so users could inspect completed actions, feasible next actions, affected dependencies, and recovery paths. 
Participants reported better understanding of task structure, step relations, and progress, and described using the graph to orient themselves before acting and to re-establish context after state changes. 
The lack of a measurable workload penalty further suggests that externalizing structure need not impose additional perceived burden when information is presented selectively. 
More broadly, a shared representation can make an AI assistant's state estimate inspectable rather than hidden.
Users can see what the system believes has happened and why it recommends a particular response. 
In this role, the graph becomes a state contract through which the user and assistant coordinate their understanding of the task.}

\textbf{Error assistance reduces users' diagnostic burden, \rw{yet it can also encourage reliance} on the assistant.}
Both the \rw{needs-finding} study and the user study suggest that proactive error assistance is valuable because users are physically and cognitively occupied during execution.
In such settings, \rw{detecting deviations, judging their significance, and formulating recovery queries for a reactive assistant becomes a second task alongside the procedure.}
\textit{RegulAR} reduced this burden by surfacing likely deviations and their consequences without waiting for user queries.
\rw{We also observed} that participants sometimes deferred their own decision-making to the assistant, especially in unfamiliar or uncertain situations.
This reliance is not inherently negative; in many cases, it reflects the usefulness of timely support.
However, it raises an important design concern: \rw{overly directive assistants} may unintentionally reduce users' engagement with their own reasoning about the task.
Future systems should therefore support error-aware guidance that assists without replacing user judgment, particularly in situations where multiple valid responses remain possible.

\textbf{Proactive assistance should be consequence-aware, layered, and user-tunable.}
Our findings also reveal that the benefits of proactivity come with a cost.
Participants appreciated timely reminders and warnings, especially when errors threatened later steps, but they did not want all deviations to trigger the same level of intervention. 
In both studies, users preferred concise notifications that first signaled that something mattered, then explained why, and finally offered recovery actions when needed. 
The user study further showed a tension between proactive assistance and autonomy: some participants valued continuous support, while others preferred more control and lighter-touch intervention. 
These findings suggest that future AR assistants should adapt not only to error type, but also to estimated downstream consequence, user expertise, and preferred level of control. 
Consequence-aware saliency, layered explanation, and mixed-initiative interaction are therefore central requirements for making proactive guidance effective without becoming overbearing.
\rc{Beyond adapting intervention saliency, future systems should also personalize the recovery strategy itself. Appropriate recovery may depend on available materials, time constraints, safety considerations, task goals, and user preferences. Rather than recommending a single generic repair, context-aware systems could generate and rank multiple valid recovery paths, allowing users to select among alternatives that prioritize factors such as quality, efficiency, appearance, or safety.}

\subsection{Limitations and Future Work}
\rw{\textbf{Evaluation remains limited in scope.}
While our evaluations provide quantitative evidence of improved procedural state tracking and qualitative evidence of better recovery support, the current evaluation remains limited in scope.}
\rc{The small sample size makes effect estimates and generalizability uncertain, and multiple comparisons may increase the risk of false positives, although significant questionnaire results remained after FDR correction.}
In particular, the study provides stronger evidence for perceived recovery support than for behavioral recovery itself. Future work should therefore \rc{include larger-scale evaluations with more diverse participant populations and directly} \rw{measure recovery outcomes, such as recovery time}, unresolved deviations, cascaded follow-on errors, and the asymmetric costs of false positives and false negatives.

\textbf{Latency, \rc{sampling}, and hallucination.}
\rc{\textit{RegulAR} performs monitoring in periodic decision cycles every five seconds. While this design aims to balance responsiveness, recognition stability, and computational cost, it may miss short-duration actions and delay the recognition of rapidly changing task states.}
\rw{In addition, the system inherits limitations from API-based multimodal models}: network latency, occasionally delayed feedback, and recognition errors or hallucinations sometimes \rw{result in incorrect assistance.}
\rw{The system also relies primarily on egocentric visual observations, leaving hidden task states} (e.g., temperature) only indirectly observable. \rw{Future work could reduce these limitations through} local inference~\cite{kulkarni_edgeai_2026}, \rc{adaptive or event-triggered monitoring strategies}, more reliable perception models~\cite{ali_hallucination_2022}, and \rw{complementary sensing modalities} such as IoT-connected devices or olfactory sensing~\cite{atzori_iot_2010,zhang_olfaction_2026}.

\rc{\textbf{Visual and cognitive load of graph-based guidance.}
Although the task graph improved progress awareness and task-structure understanding, continuously presenting graph-based guidance in AR may compete with the physical task for limited visual attention. \textit{RegulAR} mitigates this issue through impact-dependent transparency and user-controlled repositioning and resizing of the graph. Nine participants found the graph easy to understand, but two still found it distracting; one suggested registering the graph directly with relevant objects or locations in the physical environment. Future work should investigate adaptive visualization techniques that selectively reveal relevant subgraphs, reduce visual clutter, and balance global task awareness with the immediate demands of physical execution.}

\textbf{Task \rw{scalability} and long-term use remain open questions.}
Our study focused on short, single-user procedural tasks in a controlled setting.
Real-world procedures may involve longer time horizons, hidden dependencies, or collaboration, which could change both the types of errors encountered and the forms of assistance required. 
In addition, the observed autonomy trade-off suggests that intervention saliency and guidance granularity should likely adapt to users' expertise, task familiarity, and preferences for control. 
We also do not yet know how prolonged use of proactive recovery support would affect learning, trust calibration, or dependence on assistance over time. 
Future work should therefore examine whether the benefits observed here extend to longer, higher-stakes, and collaborative procedural settings, and how personalized intervention policies can better balance support and autonomy.
\rc{Scaling \textit{RegulAR} to substantially more complex procedures may introduce challenges for both system reasoning and interface presentation. Large workflows may contain hundreds of actions, deeply nested dependencies, concurrent branches, and alternative recovery paths. Existing graph-based planning and reasoning techniques could support hierarchical decomposition, subgraph retrieval, branch pruning, and incremental state updates, while visualization techniques such as semantic zooming, graph aggregation, and context-dependent disclosure could avoid presenting the full dependency structure at once. Future work should investigate how these mechanisms support large-scale and branching workflows without increasing inference latency or overwhelming users.}

\section{Conclusion}
We introduced \textit{RegulAR}, an AR task assistant for procedural error detection and recovery. By grounding MLLM-based egocentric reasoning in a hierarchical dependency graph, \textit{RegulAR} tracks task progress, recognizes errors, estimates downstream impact, and provides layered in-situ guidance that helps users understand what went wrong, why it matters, and how to proceed. Our technical evaluation and user study show that this structured approach improves objective state tracking and error recognition, while helping users build a stronger understanding of task structure, maintain progress awareness, and receive better perceived support for responding to errors. These findings suggest that the future of AR task assistance lies not only in delivering instructions but in making procedural work more recoverable when execution deviates from the plan.

\begin{acks}
\rc{The authors would like to thank all participants for their support during the studies. We are also grateful to the reviewers for their constructive feedback. Special thanks to Zhipeng Li for his insightful discussions. This work is partially supported by the Hong Kong Research Grants Council (grant\# 16214623 and T22-607/24N).}
\end{acks}

\bibliographystyle{ACM-Reference-Format}
\bibliography{main}

\appendix
\section{Study Materials}

\begin{table}[H]
  \caption{Questions used in the user evaluation.}
  \label{tab:questions}
  \centering
  \small
  \setlength{\tabcolsep}{3pt}
  \renewcommand{\arraystretch}{1.08}
  \begin{tabular}{@{}p{0.22\columnwidth}p{0.20\columnwidth}p{0.53\columnwidth}@{}}
    \toprule
    Section & Aspects & Question \\
    \midrule
    Procedural Learning
      & Structure
      & The task structure and procedure were easy to understand when using the system. \\
      & Dependency
      & The system helped me understand how different steps of the task are related. \\
      & Progress Tracking
      & The system keeps track of my progress well during the task. \\
    \midrule
    Error Assistance
      & Recognition
      & The system accurately recognized the errors I made. \\
      & Localization
      & The system helped me understand where the error occurred in the task. \\
      & Impact
      & The system helped me understand how the error affects other steps. \\
      & Urgency
      & The system helped me understand the urgency or importance of different errors. \\
      & Usefulness
      & The system provided useful guidance for error recovery. \\
      & Disruption
      & The guidance did not disrupt the completion of the task. \\
    \midrule
    System
      & Confidence
      & I successfully completed this task \\
      & Confidence
      & I would feel more confident performing this task again after using this system. \\
      & Overall
      & Overall \\
    \bottomrule
  \end{tabular}
  \Description{Questions used in the user evaluation are organized into three sections: Procedural Learning, Error Assistance, and System. Procedural Learning includes structure, dependency, and progress tracking questions. Error Assistance includes recognition, localization, impact, urgency, usefulness, and disruption. The system includes task completion confidence, future confidence, and overall rating.}
\end{table}

\begin{table}[H]
  \caption{Procedural instruction sources used in the preliminary \rw{needs-finding study (T1--T2), system testing (T3--T8), and user study (T9--T10).}}
  \label{tab:instructions}
  \centering
  \small
  \setlength{\tabcolsep}{4pt}
  \begin{tabular}{@{}p{0.08\columnwidth}p{0.49\columnwidth}p{0.12\columnwidth}@{}}
    \toprule
    ID & Task & Source \\
    \midrule
    T1  & Hybrid Meeting Setup     & \cite{markus_hybrid_meeting_video} \\
    T2  & Mug Cake                 & \cite{drummond_mugcake} \\
    T3  & Brew Coffee              & \cite{nca_pourover} \\
    T4  & Watermelon Drink         & \cite{tiffy_yakult_2024} \\
    \rc{T5}  & \rc{Elephant Toothpaste} & \cite{elephanttoothpaste} \\
    \rc{T6}  & \rc{Mac and Cheese}      & \cite{macandcheese} \\
    \rc{T7}  & \rc{Keto Pancakes}       & \cite{ketopancakes} \\
    \rc{T8}  & \rc{Italian Meatballs}   & \cite{italianmeatballs} \\
    T9  & Microwave Scrambled Eggs & \cite{pathculture_microwave_eggs_2025} \\
    T10 & Candles                  & \cite{chapman_candles_2024} \\
    \bottomrule
  \end{tabular}
  \Description{Instruction materials used across the formative study, testing, and user evaluation. T1 Hybrid Meeting Setup and T2 Mug Cake were used in the needs-finding study. T3 Brew Coffee, T4 Watermelon Drink, T5 Elephant Toothpaste, T6 Mac and Cheese, T7 Keto Pancakes and T8 Italian Meatballs were used for testing. T9 Microwave Scrambled Eggs and T10 Candles were used in the user evaluation. Each row also lists the corresponding instruction source reference.}
\end{table}

\section{Participants}

\begin{table}[H]
  \caption{Participants in the preliminary \rw{needs-finding} study. AR familiarity was rated on a 5-point scale.}
  \label{tab:formative_participants}
  \centering
  \small
  \setlength{\tabcolsep}{5pt}
  \begin{tabular*}{0.7\columnwidth}{@{\extracolsep{\fill}}llcc@{}}
    \toprule
    PID & Gender & Age & AR Familiarity \\
    \midrule
    P1 & Male   & 24 & 2 \\
    P2 & Female & 23 & 3 \\
    P3 & Male   & 25 & 3 \\
    P4 & Female & 26 & 2 \\
    P5 & Female & 23 & 2 \\
    P6 & Male   & 24 & 2 \\
    \bottomrule
  \end{tabular*}
  \Description{Demographics of needs-finding study participants, including participant ID, gender, age, and AR familiarity on a 5-point scale. Six participants were included, with three male and three female. Ages ranged from 23 to 26. AR familiarity ranged from 2 to 3 across participants.}
\end{table}

\begin{table}[H]
  \caption{\rw{Participants in the user evaluation.} \rc{AR familiarity was rated on a 5-point scale;} task familiarity was rated on a 7-point scale.}
  \label{tab:user_participants}
  \centering
  \small
  \setlength{\tabcolsep}{4pt}
  \begin{tabular*}{0.75\columnwidth}
  {@{\extracolsep{\fill}}llcccc@{}}
    \toprule
    & & & \multicolumn{3}{c}{Familiarity} \\
    \cmidrule(l){4-6}
    PID & Gender & Age & AR & T9 & T10 \\
    \midrule
    P7  & Male   & 23 & 4 & 1 & 1 \\
    P8  & Female & 23 & 4 & 5 & 2 \\
    P9  & Female & 23 & 1 & 1 & 1 \\
    P10 & Female & 22 & 2 & 2 & 2 \\
    P11 & Female & 26 & 2 & 2 & 1 \\
    P12 & Male   & 27 & 3 & 3 & 1 \\
    P13 & Female & 23 & 2 & 1 & 1 \\
    P14 & Female & 23 & 2 & 6 & 7 \\
    P15 & Male   & 21 & 1 & 6 & 2 \\
    P16 & Male   & 27 & 2 & 2 & 2 \\
    P17 & Male   & 23 & 2 & 4 & 5 \\
    P18 & Male   & 28 & 2 & 4 & 6 \\
    \bottomrule
  \end{tabular*}
  \Description{Table 5. Demographics of user evaluation participants, including participant ID, gender, age, and familiarity ratings for AR, Task 9 Microwave Scrambled Eggs, and Task 10 Candles. Twelve participants were included. Most participants reported low to moderate familiarity with AR, with ratings primarily around 2 out of 5. Task familiarity was generally low, with most participants rating both tasks 1 to 2 out of 7, although a few reported higher familiarity.}
\end{table}

\section{Prompts}

\subsection{Task Graph Generation Prompt}
\label{appendix:prompt1}

\begingroup
\scriptsize
\begin{verbatim}
This is a procedural instruction.
Output a JSON object representing a hierarchical task graph that segments
instructions into high-level Steps. The goal is to create a clean hierarchical
graph structure.

Graph Structure Rules:
1. The graph is a directed acyclic graph (DAG).
2. The graph has two node types:
   - Step nodes (high-level grouping units)
   - Action nodes (atomic executable units)
3. Each Step contains one or more Actions. Only Step "DONE" has no action
4. Steps may execute in parallel if no explicit dependency exists between them.
   Step ordering must be represented using dependencies.
5. Do NOT assume linear step order unless the text explicitly requires it.
6. Actions within the same Step must be independent and executable in parallel.
7. Do NOT split actions into separate Steps unless an explicit sequential cue
  requires it.

For each Step, include:
- step_id (unique)
- step_name (high-level summary of the grouped actions)
- step_description (ALL text related to this step that is not an atomic instruction,
  including tips, warnings, explanations, context)
- dependencies (a list of prerequisite step_ids, or null if no prerequisite exists)
- actions (list of action objects)

For each Action, include:
- action_id (unique across the entire task)
- instruction
  (single atomic sentence containing exactly ONE executable verb)
- original_text
  (the exact sentence or phrase from the source text this action was derived from)

Atomic Action Rules
- Each instruction must describe exactly ONE atomic executable action in a single
  sentence.
- Split instructions with multiple actions (e.g., "Add sugar and whisk" -> two
  separate actions).
- Split iterative actions over different materials (e.g., "Add salt, sugar, and
  vanilla extract" -> three actions).
- Do NOT hallucinate. Only use the provided information
- Only include actions that are explicit, executable instructions directly related
  to performing the task.
- Do NOT generate actions from descriptive, contextual, or explanatory text. (e.g., 
  "Add milk and serve" -> "serve" should not become an action)
- Keep every piece of relevant text in one of:
    - action.original_text
    - step_description
    - content
- Also include:
    - tools: all tools used in this action
    - materials: all materials/ingredients used in this action. If available, include
      precise amounts of the materials used.

Example Output:
{
  "steps": [
    {
      "step_id": "S1",
      "step_name": "Prepare dry ingredients",
      "step_description": "Combine the flour, salt, and baking powder.",
      "dependencies": null,
      "actions": [
        {
          "action_id": "A1",
          "instruction": "Add 1 cup of flour into the bowl.",
          "original_text": "Add 1 cup of flour into the bowl.",
          "tools":["bowl"],
          "materials": ["1 cup of flour"]
        },
        {
          "action_id": "A2",
          "instruction": "Add 1 teaspoon of salt into the bowl.",
          "original_text": "Add salt.",
          "tools": ["bowl"],
          "materials": ["1 teaspoon of salt"]
        }
      ]
    },
    {
      "step_id": "S2",
      "step_name": "Mix ingredients",
      "step_description": "Stir everything together.",
      "dependencies": ["S1"],
      "actions": [
        {
          "action_id": "A3",
          "instruction": "Mix the ingredients with a spatula.",
          "original_text": "Stir everything together.",
          "tools": ["spatula"],
          "materials": ["ingredients"]
        }
      ]
    },
    {
      "step_id": "S3",
      "step_name": "DONE",
      "step_description": "DONE",
      "dependencies": ["S2"],
      "actions": null
      }
    ],
  "content": "The estimated cook time for this piece is 5 mins. The calories are
  172kcal.",
  "summary": "Chocolate Mug Cake"
}
\end{verbatim}
\endgroup

\subsection{Task Criteria Generation Prompt}
\label{appendix:prompt2}

\begingroup
\scriptsize
\begin{verbatim}
This is information about a procedural task: "{dependency_graph}". Output a JSON
object with the following fields:

For each action, include:
- instruction (single executable sentence)
- skip_risk ("none" | "medium" | "high")
- skip_impact (textual explanation)
- status ("not_started")

- in_progress_criteria: visual indicators that the action is ongoing;
- completion_criteria: visual signs that the action is finished;
- execution_error_criteria: possible visual errors.
    - modification: corresponds to performing a step in a different way than the one
       specified by the recipe, e.g., using a different tool, such as stirring using a
        knife instead of a spoon, or using different ingredients, such as using sugar
        to sweeten the tea instead of honey. This does not necessarily change the
        outcome of the step
    - addition: corresponds to having unnecessary extra steps that are not in the task
    graph, e.g., adding raisins to the tortilla when making pinwheels
    - slip: corresponds to executing a step in a way that leads to not achieving the
     goal of the step, e.g., adding water to a different bowl from the one containing
    oats, or dropping a tortilla on the floor.

Skip Rules:
- skip_risk:
    - "none": skipping does not affect task progression and completion.
    - "medium": skipping reduces quality but does not block progression.
    - "high": skipping prevents subsequent steps from functioning.
  
Example Outputs: 
{
  "steps": [
    {
      "step_id": "S1",
      "step_name": "Prepare dry ingredients",
      "step_description": "Combine the flour, salt, and baking powder.",
      "dependencies": null,
      "actions": [
        {
          "action_id": "A1",
          "instruction": "Add 1 cup of flour into the bowl.",
          "original_text": "Add 1 cup of flour into the bowl.",
          "tools": ["mixing bowl", "measuring cup"],
          "materials": ["1 cup flour"],
          "skip_risk": "high",
           "skip_impact": "Without flour, the mixture cannot form, and later
           steps cannot proceed.",
          "status": "not_started",
          "in_progress_criteria": "Flour is visibly pouring into the bowl.",
           "completion_criteria": "Measured flour is inside the bowl, and the
           cup is empty.",
          "execution_error_criteria": {
            "modification": "Using a different ingredient instead of flour.",
            "addition": "Adding extra dry ingredients not specified.",
            "slip": "Pouring flour outside the bowl."
          }
        },
        {
          "action_id": "A2",
          "instruction": "Add 1 teaspoon of salt into the bowl.",
          "original_text": "Add salt.",
          "tools": ["mixing bowl", "measuring spoon"],
          "materials": ["1 teaspoon salt"],
          "skip_risk": "medium",
          "skip_impact": "Skipping salt reduces flavor quality but does not
          block progression.",
          "status": "not_started",
          "in_progress_criteria": "Salt is being measured or poured into
          the bowl.",
          "completion_criteria": "Measured salt is visibly inside the bowl.",
          "execution_error_criteria": {
            "modification": "Using sugar instead of salt.",
            "addition": "Adding extra seasoning is not required.",
            "slip": "Adding excessive salt due to mis-measurement."
          }
        }
      ]
    },
    {
      "step_id": "S2",
      "step_name": "Mix ingredients",
      "step_description": "Stir everything together.",
      "dependencies": ["S1"],
      "actions": [
        {
          "action_id": "A3",
          "instruction": "Mix the ingredients until the mixture looks uniform.",
          "original_text": "Stir everything together.",
          "tools": ["spatula", "mixing bowl"],
          "materials": ["combined ingredients"],
          "skip_risk": "high",
          "skip_impact": "If not mixed, ingredients remain uneven, and the batter
          is incomplete.",
          "status": "not_started",
          "in_progress_criteria": "The spatula is moving, and the mixture texture
          is changing.",
          "completion_criteria": "No visible dry powder remains, and texture
          is consistent.",
          "execution_error_criteria": {
            "modification": "Using a different tool such as a fork.",
            "addition": "Adding new ingredients during mixing.",
            "slip": "Stopping too early and leaving visible lumps."
          }
        }
      ]
    },
    {
      "step_id": "S3",
      "step_name": "Optional decoration",
      "step_description": "Add powdered sugar on top.",
      "dependencies": ["S2"],
      "actions": [
        {
          "action_id": "A4",
          "instruction": "Sprinkle powdered sugar on top.",
          "original_text": "Dust with powdered sugar.",
          "tools": ["sifter"],
          "materials": ["powdered sugar"],
          "skip_risk": "none",
          "skip_impact": "Skipping this action does not affect task completion;
          it only changes appearance.",
          "status": "not_started",
          "in_progress_criteria": "Powdered sugar is falling onto the surface.",
          "completion_criteria": "A thin visible layer covers the top.",
          "execution_error_criteria": {
            "modification": "Using a different topping.",
            "addition": "Adding extra toppings not specified.",
            "slip": "Dumping too much sugar in one spot."
          }
        }
      ]
    },
    {
      "step_id": "S4",
      "step_name": "DONE",
      "step_description": "Task completed.",
      "dependencies": ["S3"],
      "actions": null
    }
  ],
  "content": "The estimated cook time for this piece is 5 mins. The calories are
  172kcal.",
  "summary": "Chocolate Mug Cake"
}
\end{verbatim}
\endgroup

\end{document}